\documentclass{article}
\usepackage{graphicx}
\usepackage{color}
\def\beq{\begin{equation}}
\def\eeq{\end{equation}}
\usepackage{graphicx}
\usepackage{amsmath}
\usepackage{amsfonts}
\usepackage{amssymb}
\usepackage{amscd}
\def\pf{\tilde{\mathbf{p}}}
  
\def\xf{\tilde{\mathbf{x}}}
\def\yf{\tilde{\mathbf{y}}}
\def\ff{\tilde{{f}}}
\def\gf{\tilde{{g}}}

\begin{document}

\begin{center}
{\bf The light-front vacuum
}
\end{center}

\begin{center}
{
W. N. Polyzou \\
The University of Iowa \\
Marc Herrmann\\
The University of Iowa 
}
\end{center}

\noindent{\bf abstract:} We discuss the relation between the trivial 
light-front vacuum and the non-trivial Heisenberg vacuum. 

\section{Introduction}
\label{intro}

The light-front representation \cite{Dirac} of quantum field theory
\cite{Chang} has the
advantage that the vacuum of the interacting theory is identical to
the Fock vacuum of the corresponding free field theory.  This is in
contrast to the canonical representation of field theory where the
interacting vacuum has infinite norm in the free field Hilbert space.
On the other hand the light-front and canonical representations of the
same field theory should be equivalent.  The purpose of this work is
to understand the relation between the vacuum vectors in these two
representations of the same field theory.

\section{Triviality of the light-front vacuum}
\label{triviality}

The light front is a hyperplane tangent to the light cone
defined by  $x^+= x^0 + \hat{\mathbf{n}}\cdot \mathbf{x}=0$.
Translations tangent to the light-front hyperplane 
are independent of interactions and are generated by the following 
components of the four momentum
\begin{equation}
P^{+} := P^0 + \hat{\mathbf{n}}\cdot \mathbf{P} \geq 0
\qquad
\mathbf{P}_{\perp} := \mathbf{P} - \hat{\mathbf{n}}
(\hat{\mathbf{n}}\cdot \mathbf{P}) .
\label{eq.1}
\end{equation}
Interactions, $V$, appear in the generator of translations
normal to the light-front hyperplane
\begin{equation}
P^- = P_0^- + V.
\label{eq.2}
\end{equation}
The translation generators $P^+$ and $\mathbf{P}_{\perp}$ are sums of
single-particle operators
\begin{equation}
P^+ = \sum_n p_n^+; \qquad \mathbf{P}_{\perp} = \sum_n \mathbf{p}_{n\perp},
\label{eq.3}
\end{equation}
where each $p_n^+ \geq 0$. 

The origin of the triviality of the light-front vacuum follows because
both $P^-$ and $P^-_0$ commute with $P^+$.  This means that 
$V \vert 0 \rangle$, where $\vert 0 \rangle$ is the free Fock vacuum,
is an eigenstate of $P^+$ with eigenvalue
zero.  If $V$ is expressed in normal ordered form,
the part of $V$ with all creation operators can only increase the
value of $p_n^+$
\begin{equation}
V =  \int {\cal V} (\pf_1, \cdots ,\pf_n ) a^{\dagger} (\pf_1) \cdots
a^{\dagger} (\pf_n) d\pf_1 \cdots d\pf_n + \cdots , 
\label{eq.4}
\end{equation}
however the sum of all of the non-negative $p_n^+$  must vanish, so
it follows that either ${\cal V} \vert 0 \rangle =0$ or the pure
creation operator coefficient is proportional to $\delta (p^+_n)$ or
derivatives of $\delta (p^+_n)$.  We refer to these contributions, if
they appear, as zero modes.  These contributions cannot appear if the
kernel ${\cal V}$ is smooth.  In this case the
interaction does not change the Fock vacuum.  This does not happen in
the canonical case because although the interaction is translationally
invariant, the single-particle momenta can sum to zero without being
all zero.

\section{Characterization of the vacuum}
\label{characterization}

The simplest way to understand the relation between the light-front
and canonical vacuua is to first consider the case of free fields,
where the precise relations between the Heisenberg field, fields
restricted to a fixed time surface and fields restricted to a light
front are known.  For a free scalar field the fixed time and light
front representations of the field are given by the well-known
expressions:
\begin{equation}
\phi (x) = {1 \over (2 \pi)^{3/2}}
\int {d \mathbf{p} \over \sqrt{2 \omega_m(\mathbf{p})}}
(e^{i p\cdot x} a(\mathbf{p}) + e^{-i p\cdot x} a^{\dagger}(\mathbf{p})) 
\label{eq.5}
\end{equation}
\begin{equation}
={1 \over (2 \pi)^{3/2}}
\int {d \pf \theta (p^+) \over \sqrt{2 p^+}}
(e^{i p\cdot x} a(\pf) + e^{-i p\cdot x} a^{\dagger}(\pf)), 
\label{eq.6}
\end{equation}
where $\omega_m(\mathbf{p})$ is the energy of a particle of mass
$m$ and momentum $\mathbf{p}$ and $\pf = (p^+,\mathbf{p}_{\perp})$. 

In either representation the vacuum is normally defined by the
condition that it is annihilated by the annihilation operator.
Identifying these two representations of the field and changing
variables from the three momenta to the light-front components of the
four momentum gives the relation between the annihilation operators
\begin{equation}
a_i (\mathbf{p}) = a_i (\pf)\sqrt{p^+ \over \omega_{m_i}(\mathbf{p})}.
\label{eq.7}
\end{equation}
This suggest that the vacuum defined by the annihilation operator in
these two representations is the same.  When the field is restricted
to the light front all of the mass dependent (dynamical) terms in
(\ref{eq.6}) disappear.  Furthermore, the field restricted to the
light front is irreducible (i.e. it is possible to extract both the
creation and annihilation operator directly from the field restricted
to the light front - there is no need for derivatives off of the light
front).  It follows that all inner products and all expectation values
of operators are identical, which means that free fields with
different masses restricted to the light front are unitarily
equivalent. This suggests that there is one vacuum vector, up to
unitary equivalence, for free scalar fields of any mass.  On the other
hand, the creation and annihilation operators for two free scalar
fields with different masses are related by the canonical
transformation \cite{haag}
\begin{equation}
a_2 (\mathbf{p}) = \cosh (\eta (\mathbf{p}) )  a_1 (\mathbf{p}) + 
\sinh (\eta (\mathbf{p})) a_1^{\dagger} (\mathbf{p})
\label{eq.8}
\end{equation}
where 
\begin{equation}
{\cosh (\eta (\mathbf{p}) )} = 
{1 \over 2} \left (\sqrt{{\omega_{m_2}(\mathbf{p})\over
    \omega_{m_1}(\mathbf{p})}}
+ \sqrt{{\omega_{m_1}(\mathbf{p})\over \omega_{m_2}(\mathbf{p})}}\right ).
\label{eq.9}
\end{equation}
The problem is that the generator of the associated unitary transformation 
\begin{equation}
G= 
-i \int { d\mathbf{p}\eta (\mathbf{p})
\over 2 }\left (a_1(\mathbf{p})a_1(\mathbf{p})-
a^{\dagger}_1(\mathbf{p})a^{\dagger}_1(\mathbf{p})\right )
\label{eq.10}
\end{equation}
has infinite norm when applied to the vacuum of particle $1$, which 
means the canonical transformation (\ref{eq.8}) cannot be realized as a unitary 
transformation on the Hilbert space associated with particle 1.  

These results have the appearance of being inconsistent - specifically
identifying the vacuum of the canonical and light-front fields of the
same mass, and the vacuum of the light-front fields with different
masses, suggest that all of the vacuua are identical or related by a
unitary transformation.  On the other hand a direct comparison on the vacuua
of canonical field theories with different masses show that they are
not even in the same Hilbert space .

The resolution of this apparent inconsistency is that the annihilation
operator does not completely characterize the vacuum \cite{marc}.  Another
characterization of the vacuum is as an invariant linear functional on
an algebra of operators.  The relevant algebras are the algebra of 
free fields integrated against test functions of four space-time
variables, the algebra of free fields restricted to a light front,
integrated against test functions in coordinates of the light-front
hypersurface, and the algebra of free fields and their time derivatives
restricted to a fixed time surface,  integrated against test functions
in three space variables.  We call these algebras the local algebra,
the light-front algebra, and the canonical algebra respectively.
Each algebra is invariant with respect to a different symmetry group.

The formal expressions (\ref{eq.5}-\ref{eq.6}) of the fields in terms
of creation and annihilation operators can still be used to construct
elements of each of these algebras by making the appropriate
restrictions and integrating against the appropriate test functions.

From the algebraic perspective it is clear that both the light-front and
canonical algebras do not contain enough operators to formulate
Lorentz invariance or locality.  Schlieder and Seiler \cite{Schlieder} give
an example that illustrates the importance of the underlying algebra.
They consider two free fields of different mass, but they restrict the
algebra by limiting the test functions to functions with Fourier
transforms that are related on the two different mass shells by
\begin{equation}
{f (\sqrt{m_1^2 + \mathbf{p}^2},\mathbf{p}) \over   
(m_1^2 + \mathbf{p}^2)^{1/4}} =
{f (\sqrt{m_2^2 + \mathbf{p}^2},\mathbf{p}) \over   
(m_2^2 + \mathbf{p}^2)^{1/4}}.
\label{eq.11}
\end{equation}
For this class of test functions 
\begin{equation}
\langle 0_1 \vert \phi_1 (f_1) \cdots \phi_1 (f_n) \vert 0_1\rangle=
\langle 0_2 \vert \phi_2 (f_1) \cdots \phi_2 (f_n) \vert 0_2\rangle .
\label{eq.12}
\end{equation}
In this case the inner products are identical, and the two fields
restricted to this algebra are related by a unitary transformation.
In addition, because there is no restriction on the test functions on
one mass shell, it follows that both of these algebras are
irreducible.  On the other hand, these properties are not preserved
when the restricted class of test functions are enlarged to include the full
set of test functions.  This illustrates the importance of
the role of the algebra.

In the case of the two representations of the free fields, the
structure of the field (\ref{eq.5}-\ref{eq.6}) plays a key role in
extending the algebra from an algebra of operators restricted to a
hypersurface to the local algebra.  In the case of the light-front
representation the free Heisenberg field of mass $m$ can be expressed
as a linear operator acting on a field restricted to a light front
\begin{equation}
\phi (x) = \int F_m (x,\yf) \phi (\yf) d\yf 
\label{eq.13}
\end{equation}
where 
\begin{equation}
F_m (x,\yf) =
{1 \over (2 \pi)^2} 
\int {d\pf \over 2} 
e^{-i {p^2_\perp + m^2 \over p^+}x^+ + i \pf \cdot (\xf-\yf)}.
\label{eq.14}
\end{equation}
Here the kernel $F_m (x,\yf)$ that defines this extension has
the mass dependence, which is the dynamical information for a free
field.  We can also alternatively interpret this kernel as
a map of test functions in
four variables to test functions restricted to a light front.
The Fourier transforms of these test
functions are related by 
\begin{equation}
\ff (\pf) = f \left ( {\mathbf{p}_{\perp}^2 + m^2 \over p^+},\pf
\right ) . 
\label{eq.15}
\end{equation}
Here we use a notation where a $\ff$ indicates a function of
light-front variables.  The important observation is that because the
Fourier transform of a Schwartz function in four variables is a
Schwartz function, the function $\ff (\pf)$ vanishes faster than
$(p^+)^n$ for any $n$ at the origin.  We will see that this
observation has important implications when we discuss properties of
the light-front Fock algebra.

We can interpret $F_m (x,\yf)$ as a mapping from the local Heisenberg
algebra to a sub-algebra of the light-front Fock algebra that
preserves all Wightman functions, and is hence unitary.
Under this mapping the vacuum becomes the free field Fock vacuum,
and the dynamics is contained in the mapping.  

\section{Zero modes and dynamics}
\label{zeromodes}

In order to consider the potential role of zero modes it is useful to
summarize the key properties of the light-front Fock algebra \cite{leutwyler}\cite{Schlieder}\cite{coester}, that is that
algebra generated by free fields restricted to a light front.
Operators in this algebra have the form 
\begin{equation}
\phi (\ff)= \int d \xf \phi (\xf,x^+=0) \ff (\xf) .
\label{eq.16}
\end{equation}
This algebra is closed under operator multiplication, which
can be summarized by 
\begin{equation}
e^{i \phi (\ff) } e^{i \phi (\gf)}
= e^{i \phi (\ff+\gf)} e^{-{1\over 2} ((\ff,\gf)-(\gf,\ff))}
\label{eq.17}
\end{equation}
where the light-front scalar product is given by
\begin{equation}
(\ff,\gf) = \int {d\pf \theta (p^+) \over p^+}
\ff(-\pf) \gf(\pf) .
\label{eq.18}
\end{equation}
One of the important properties of this algebra is that it is
irreducible.  The simplest way to understand this is to note that it
is possible to extract the creation and annihilation parts of $\phi
(\ff)$ without extending the algebra.  Specifically the creation and
annihilation operators are related to the Fourier transform of the
restricted field by
\begin{equation}
\phi_+ (\pf):= \theta (p^+) \phi (\pf) = {\theta (p^+) \over \sqrt{p^+}}a(\pf) 
\label{eq.19}
\end{equation}
\begin{equation}
\phi_- (\pf):= \theta (-p^+) \phi (\pf) = {\theta (-p^+) \over \sqrt{-p^+}}
a^{\dagger} (-\pf) 
\label{eq.20}
\end{equation}
\begin{equation}
\phi (\pf) = \phi_+ (\pf) + \phi_- (\pf) .
\label{eq.21}
\end{equation}
One consequence of this is that there is a purely algebraic
normal ordering, which can be summarized by
\begin{equation}
:e^{i \phi (\ff) }: = e^{i \phi_-(\ff)} e^{i\phi_+(\ff)} .
\label{eq.22}
\end{equation}
It follows, using the Campbell-Baker-Hausdorff formula that
\begin{equation}
e^{i \phi (\ff) } = :e^{i \phi(\ff)}: e^{{1 \over 2}(\ff,\ff)} .
\label{eq.23}
\end{equation}

We note that independent of the choice of vacuum, if the test function
vanishes for $p^+=0$, the light-front inner product in (\ref{eq.23}) is
well behaved and the vacuum expectation value of the normal product
must be one, since for $p^+\not=0$ the annihilation operator
necessarily reduces the value of $p^+$, which means that it must
annihilate the vacuum.  In this case the vacuum is uniquely
determined by this equation.  Since the operator $F_m$ maps four
variable test functions into functions on the light front with Fourier
transforms that vanish for $p^+=0$, the vacuum is always the free
field Fock vacuum in this case and there can be no zero mode
contributions.

It is well known that the equivalence of calculations of some
observables based on light-front field theory and canonical field
theory require 0-mode contributions in the light-front expressions.
So far we have only considered free fields, so there is a question
concerning whether zero modes are a consequence of a more complex
dynamics or not.

To answer this first note that the classical Noether's theorem  
gives formal expressions for the Poincar\'e generators 
as local products of operators at the same point on the light front.
These products are not elements of the light-front
algebra.  One can see the problem in the following expression for the
two point Wightman function in light-front variables:
\begin{equation}
\langle 0 \vert \phi (x) \phi (y) \vert 0 \rangle =
\label{eq.24}
\end{equation}
\begin{equation}
{1 \over (2 \pi)^3} \int {d\pf \over {p^+}} 
e^{-i {\mathbf{p}_{\perp}^2 + m^2 \over {p^+}}(x^+-y^+) 
+i \pf (\xf - \yf)} .
\label{eq.25}
\end{equation}
Although there is an apparent logarithmic singularity due to the
$p^+$ integral, if $(x^+-y^+)$ is not zero
the $1/p^+$ singularity in the above expression is regulated
by oscillations in the exponent.  This can be 
seen by considering the integral of the same form
\begin{equation}
\int_0^a {e^{i c/p^+} \over p^+ } dp^+ =
\int_{c/a}^\infty {e^{iu }\over u } du = {\pi \over 2} -(Ci(c/a)+ i Si
(c/a)) .
\label{eq.26}
\end{equation}

If both fields are restricted to the light front, this mechanism that
regulates the singularity at $p^+=0$ is turned off leading an infrared
singularity.  There will also be ultraviolet singularities associated
with the other variables.  In this case there is no reason for the
mapped test functions to vanish at $p^+=0$.  In this case if we try to
make sense out of (\ref{eq.25}) when the test function does not vanish at
$p^+=0$ two things happen.  First the logarithmic singularity in the
light front scalar product needs to be regularized.  Typically this
regularization breaks longitudinal boost invariance.  A second thing
that can happen is that because it is possible to have non-zero
contributions from $p^+=0$ the algebraic normal product can be
extended to include additional singular contributions that have
support on $p^+=0$.  These can have a non-trivial dependence on
transverse momentum and may be needed to restore full rotational
covariance and longitudinal boost invariance.

It follows that zero modes are a consequence of renormalizing 
local operator
products.  They are available tools that may be needed to make the
theory finite and consistent with the corresponding canonical theory.

We note that Lorentz invariance of the $S$-matrix in a light-front quantum
theory is equivalent to invariance with respect to a change of orientation
of the light front \cite{polyzou}.  Changing the orientation of 
the light front can
exchange infrared and ultraviolet singularities. This suggest that
the problem of zero modes and renormalization is more complicated in
3+1 dimensional theories than it is in 1+1 dimensional theories.

While the above analysis is limited to free fields,
in an asymptotically complete field theory an interacting
Heisenberg field can be expanded in normal ordered products of an
irreducible set of asymptotic (in or out) fields\cite{haag}\cite{Glaser}:
\begin{equation}
\phi (x) = \sum \int {L}(x;z_1, \cdots , z_n) {:\phi_{in} (z_1) 
\cdots \phi_{in} (z_n):}dz_1 \cdots dz_n . 
\label{eq.27}
\end{equation}
The asymptotic fields are free fields, and following
the above analysis, can be expressed as products of free fields
on the light front using the mapping (\ref{eq.14}):
\begin{equation}
\phi (x) = \sum \int {\cal L}(x;\yf_1, \cdots , \yf_n) {:\phi_0 (\yf_1) 
\cdots \phi_0 (\yf_n):}d\yf_1 \cdots d\yf_n 
\label{eq.28}
\end{equation}
\begin{equation}
{\cal L}(x;\yf_1, \cdots , \yf_n) = 
\int L(x;z_1, \cdots ,z_n) \prod_i 
F_m(z_i ,\yf_i )d^4 z_1 \cdots d^4z_n . 
\label{eq.29}
\end{equation}
Furthermore, if the Heisenberg field $\phi (x)$ is an operator valued
distribution we expect that the smeared coefficient
will behave like (\ref{eq.15}):
\begin{equation}
\int f(x) \tilde{{\cal L}}(x;\pf_1, \cdots , \pf_n)d^4 x \to 0
\quad \mbox{any} \quad p_i^+ \to 0 .
\label{eq.30}
\end{equation}
If this is the case then $\{ {\cal L}(x;\yf_1, \cdots ,\yf_n) \}$
is the kernel of a unitary mapping from the Heisenberg algebra of 
the local field
field theory to a sub-algebra of the light-front Fock algebra, where the
unitary transformation maps the Heisenberg vacuum to the trivial Fock
vacuum:
\begin{equation}
\phi (x) = \sum \int {\cal L}(x;\yf_1, \cdots ,\yf_n) :\phi_0 (\yf_1) 
\cdots \phi_0 (\yf_n):d\yf_1 \cdots d\yf_n .
\label{eq.31}
\end{equation}
At this level zero modes cannot occur, however they can
occur as intermediate steps in the construction of the
coefficient functions
$\{ {\cal L}(x;\yf_1, \cdots ,\yf_n) \}$.

The conclusion of this work is that there is a unitarily equivalent
representation of the dynamics with the free-field vacuum.  The
Hilbert space is generated from this vacuum by a sub-algebra of the
light-front Fock algebra.  While the vacuum functional has no zero
mode contributions on this sub algebra, zero modes can appear 
in intermediate steps in the
construction of the operators that map the local Heisenberg algebra to
this sub-algebra.

\end{document}